\documentstyle[prl,aps,epsf]{revtex} \def\narrowtext{} \tighten \twocolumn
\input epsf.sty 
\begin{document}
\draft
 
\title{Phenomenology of Photoemission Lineshapes 
of High Tc Superconductors }
\author{
        M. R. Norman,$^1$
        M. Randeria,$^2$
        H. Ding,$^{1,3}$
        and J. C. Campuzano,$^{1,3}$
       }
\address{
         (1) Materials Sciences Division, Argonne National Laboratory,
             Argonne, IL 60439 \\
         (2) Tata Institute of Fundamental Research, Mumbai 400005, India\\
         (3) Department of Physics, University of Illinois at Chicago,
             Chicago, IL 60607\\
         }

\address{%
\begin{minipage}[t]{6.0in}
\begin{abstract}
We introduce a simple phenomenological form for the self-energy 
which allows us to extract important information from angle resolved 
photoemission data on the high Tc superconductor Bi2212. 
First, we find a rapid suppression of the single particle scattering rate
below $T_c$ for {\it all} doping levels.
Second, we find that in the overdoped materials the gap $\Delta$ at
all {\bf k}-points on the Fermi surface has significant temperature 
dependence and vanishes near $T_c$. In contrast, 
in the underdoped samples such behavior is found only at {\bf k}-points
close to the diagonal. Near $(\pi,0)$, $\Delta$ is essentially 
$T$-independent in the underdoped samples. The filling-in of 
the pseudogap with increasing $T$ is described by a broadening 
proportional to $T-T_c$, which is
naturally explained by pairing correlations above $T_c$.
\typeout{polish abstract}
\end{abstract}
\end{minipage}}

\maketitle
\narrowtext

Angle-resolved photoemission spectroscopy (ARPES) has played a major role in 
developing our understanding of the high $T_c$ superconductors.
The momentum and frequency resolved information contained in the
one-particle spectral function \cite{NK} probed by ARPES
provides critical insights difficult to obtain from other techniques.
However, an important open problem in interpreting ARPES
data is the absence of simple representations
of the spectral lineshape, analogous to the Drude formula for optical
studies. 

In this paper, we take a step in this
direction by introducing a simple phenomenological form for the self-energy 
which captures much of the important low frequency information contained
in the ARPES data for the normal, superconducting and pseudogap phases
of high $T_c$ superconductors.  
We first test this on ARPES data at ${\bf k}_F$ for overdoped Bi2212,
from which we determine the $T$-dependence of the superconducting (SC)
gap and the one-particle scattering rate, the latter found to decrease
rapidly in the SC state, leading to the appearence of sharp quasiparticles
at low temperatures. 

We then turn to underdoped Bi2212, 
where a highly anisotropic pseudogap is known to persist
well above $T_c$ \cite{STANFORD,DING,PRL97}. The magnitude of this gap
and its smooth evolution through $T_c$ establish a strong
connection between the normal state pseudogap and the SC gap below $T_c$,
and suggest that the pseudogap arises from pairing correlations in a state
without long range phase coherence \cite{REVIEW}.
Here, we use our simple self energy to gain insight into the ARPES 
lineshape in the pseudogap regime. We find that near $(\pi,0)$
$\Delta$ is $T$-independent, and the pseudogap fills in due to a 
$T$-dependent broadening which is naturally explained by pairing 
correlations above $T_c$.

Finally, we use our analysis to shed more light on the
very recent surprising result that the pseudogap 
at different {\bf k}-points turns off at different temperatures,
leading to gapless arcs above $T_c$ which expand in length until the 
entire Fermi surface is recovered at $T^*$ \cite{NEW}.
We show below that the $T$-dependence of the underdoped lineshape
away from $(\pi,0)$ is rather different, with
the vanishing of the pseudogap
controlled instead by a $T$-dependent $\Delta$.

The data analysed below have been reported earlier \cite{DING,PRL97,NEW}.  
The ARPES intensity $I({\bf k},\omega)$ is proportional
to $f(\omega)A({\bf k},\omega)$, where $f$ is the Fermi function
and $A$ the spectral function \cite{NK}. 
Recently, we have proposed a method \cite{NEW} which allows us to eliminate 
the effect of $f$ from ARPES data and focus directly on $A$. 
In brief, using the mild assumption of particle-hole symmetry
$A(-\epsilon_{\bf k},-\omega) = A(\epsilon_{\bf k},\omega)$ for small
$|\omega|$ and within the small ${\bf k}$-window centered
at ${\bf k}_F$, one can show that the symmetrized intensity
$I(\omega) + I(-\omega)$ at ${\bf k}_F$ is simply the
spectral function (convolved with the resolution) \cite{FOOT1}.
Results obtained from symmetrized data agree with those 
obtained from the leading edge of the raw data \cite{NEW}.  
Here, we use the symmetrized data for two reasons: it is a
useful visual aid, and (due to absence of Fermi cutoff)
it allows more stringent comparisons to the fits. 
We have checked that the fits discussed below agree equally well with 
the raw data.

We begin with the overdoped samples, where there are no
strong pseudogap effects. The simplest
self-energy which can describe the data at all $T$ is
\begin{equation}
\Sigma({\bf k},\omega) = -i\Gamma_1
+ \Delta^2/[(\omega +i0^+) + \epsilon({\bf k})].
\label{SE.1}
\end{equation}
Here $\Gamma_1$ is a single-particle scattering rate
taken, for simplicity, to be an $\omega$-independent constant.
It is effectively an average of the (actual $\omega$-dependent)
$\Sigma''$ over the frequency range of the fit.
The second term is the BCS self-energy 
(corresponding to the diagonal term of the Nambu-Gorkov propagator) 
where $\Delta$ is the SC gap, 
and $\epsilon({\bf k})$ the dispersion (with $\epsilon({\bf k}_F)=0$). 
We emphasize that this is a minimal model; modeling the full
$\omega$-dependence of the self-energy will involve more than 
$\Gamma_1$ and $\Delta$.

In Fig.~1a, we show symmetrized data for an overdoped $T_c$=82K sample 
at ${\bf k}_F$ along $(\pi,0)-(\pi,\pi)$ together with
the fits obtained as follows. 
Using Eq.~1 we calculate the spectral function
$\pi A({\bf k},\omega) = \Sigma''({\bf k},\omega) / 
\left[(\omega - \epsilon_{\bf k} - \Sigma'({\bf k},\omega))^2
+ \Sigma''({\bf k},\omega)^2\right]$,
convolve it with the experimental resolution, and fit to 
symmetrized data.  The fit is restricted to a range of $\pm$45 meV 
given the small gap in the overdoped case and the sharpness of the 
quasiparticle peaks below $T_c$. We find that Eq.~1
describes the data quite well\cite{HIGH}. 

The $T$-variation of the fit parameters
$\Delta$ and $\Gamma_1$ are shown in Fig.~2a.
$\Delta (T)$ decreases with $T$, and although small at $T_c$, it only vanishes
above $T_c$, indicating the possibility of a weak pseudogap.  This effect is
sample dependent, in that several overdoped samples we have looked at, the gap
vanishes closer to $T_c$.  We caution that the error bars shown in Fig.~2a
are based on the RMS error of the fits, but do not take into account
experimental errors in $\mu$ and ${\bf k}_F$.

$\Gamma_1(T)$ is found to be
relatively $T$-independent in the normal state.
In comparing this to optical conductivity data,
it must be kept in mind that $\Gamma_1$ is 
an average of the true $\Sigma''$ near $(\pi,0)$
over the $\omega$-range of the fit, whereas $1/\tau_{tr}$
is a zone average weighted by velocity
factors which also differs from $\Sigma''$ due
to different thermal factors.  Despite this, it is interesting
to note that optical conductivity work has also found a relative lack of
T dependence to $1/\tau_{tr}(\omega)$ for $\omega>2\Delta$\cite{PUCH}.
Below $T_c$,
we see that $\Gamma_1$ decreases very rapidly, and 
can be perfectly fit to the form $a + bT^6$ \cite{RESIDUAL}.  
This rapid drop in linewidth leading to sharp quasiparticle peaks
at low $T$, which can be seen directly in the ARPES data,
is consistent with SC state microwave and thermal conductivity 
\cite{TRANSPORT} measurements, and implies that electron-electron 
interactions are responsible for $\Gamma_1$ \cite{NK,HIGH}.
Note the clear break in $\Gamma_1$ at $T_c$, despite the fact $\Delta$ has not
quite vanished.
We have seen similar behavior to that described above
for a variety of overdoped samples at several {\bf k} points.

We next turn to the more interesting underdoped case.  We find 
that near $(\pi,0)$ the self-energy (1) cannot give an adequate
description of the data, in that it does not
properly describe the pseudogap and its unusual ``filling in" above $T_c$. 
Theoretically, we cannot have a divergence in $\Sigma({\bf k}_F,\omega=0)$
in a state without broken symmetry.
A simple modification of the BCS self-energy rectifies both these problems:
\begin{equation}
\Sigma({\bf k},\omega) = -i\Gamma_1 + \Delta^2 /
[\omega + \epsilon({\bf k}) + i\Gamma_0].
\label{SE.2}
\end{equation}
The new term $\Gamma_0 (T)$ should be viewed as the inverse pair lifetime;
below $T_c$, where the pairs have infinite lifetime, 
$\Gamma_0=0$, and (\ref{SE.2}) reduces to (\ref{SE.1}).
The theoretical motivation for (\ref{SE.2}) is given in the Appendix.
We stress that this three parameter form is again a minimal representation 
of the pseudogap self-energy. Since it is not obviously a unique 
representation, it is very important to see what one learns from the fits.

In Fig.~1b, we show symmetrized data at the $(\pi,0)-(\pi,\pi)$ Fermi 
crossing for a $T_c$=83K underdoped sample. 
Below $T_c$ we see quasiparticle peaks. Above $T_c$ these 
peaks disappear but there is still a large suppression of spectral
weight around $\omega$=0.  As $T$ is raised further, the pseudogap fills in
(rather than 
closing) leading to a flat spectrum at a temperature of $T^*$ (200K). 
The self-energy (\ref{SE.2}) gives a good fit to the data.  These
fits were done below $T_c$ over a larger energy range ($\pm$75 meV) than in
the overdoped case because of the larger SC gap.
The range above $T_c$ was increased to $\pm$85 meV so as to
properly describe the pseudogap depression.

In Fig.~2b, we show the $T$-dependence of the fit parameters.
We find a number of surprises.  First, $\Delta$
is independent of $T$ within error bars.  
Similar behavior has been infered from specific heat\cite{LORAM}
and tunneling\cite{RENNER} data.
This $T$-independence is in total contrast to the behavior of 
the overdoped 82K sample with almost identical $T_c$ at the same {\bf k} 
point. In addition, for the underdoped sample,
the gap evolves smoothly through $T_c$.  

The single-particle scattering rate $\Gamma_1(T)$ for the
underdoped sample is found to be qualitatively similar to the overdoped 
case. It is consistent with being
$T$-independent above $T_c$, but with a value over twice as large as the
overdoped case (allowing $\Gamma_1$ to vary above $T_c$ does not improve the
RMS error of the fits). 
Second, we see the same rapid decrease in $\Gamma_1$
below $T_c$ as in the overdoped case\cite{RESIDUAL}.
Note again the clear break at $T_c$.

The most interesting result is $\Gamma_0(T)$.
We find $\Gamma_0 = 0$ below $T_c$ and proportional to $T-T_c$ above.  
This behavior is robust, and is seen in all the fits that we have tried.
Moreover, a non-zero $\Gamma_0$ is needed above $T_c$ to obtain a proper fit
to the data (its effect cannot be reproduced by varying the other
parameters).
The fact that this $T$-dependence is exactly
what one expects of an inverse pair lifetime
(with a prefactor about twice the weak coupling BCS value;
see Appendix),
is a non-trivial check on the validity of the
physics underlying Eq.~2.
Further, we observe from Fig.~2 that $T^*$ corresponds to
where $\Delta(T) \sim \Gamma_0(T)$.  This condition can be understood from
the small $\omega$ expansion of Eq.~2.

We note, in passing, that both ARPES \cite{HARRIS,SNS97} and tunneling
measurements\cite{MIYA} indicate that the low temperature SC gap $\Delta(0)$
in Bi2212 increases as the doping is reduced. This increase closely tracks
that of the $T^*$ values determined from ARPES \cite{SNS97}.  
This is further evidence linking $T^*$ to the onset of pairing.

The next important question is whether the $T$-dependence near $(\pi,0)$
described above exists at other ${\bf k}_F$-points. 
To answer this, we have looked at 
$T$-dependent data for a number of underdoped samples at two different 
{\bf k} vectors\cite{NEW}.  All data at the $(\pi,0)-(\pi,\pi)$ 
Fermi crossing give results similar to those for the 83K sample.  
However at the second {\bf k}-point, about
halfway between the first and the d-wave node along the $(0,0)-(\pi,\pi)$ 
direction, we see quite different behavior.  
We demonstrate this in Fig.~3a where symmetrized data for a 77K 
underdoped sample are shown. For the first {\bf k} point, 
one clearly sees the gap fill in above $T_c$, with little evidence 
for any $T$-dependence of the position of the spectral feature 
defining the gap edge, just as for the 83K sample.  
In contrast, at the second {\bf k} point, the gap is clearly closing, 
indicating a strong $T$-dependence of $\Delta$. 
Similar behavior is seen in other underdoped samples with
$T_c$ between 75 and 85K.  

In Fig.~3b, we show the $T$-dependence of
$\Delta$ obtained from fits (over a range of $\pm$66 meV)
at the second {\bf k} point for the 77K sample.
$\Delta$ is found to be strongly 
$T$-dependent, being roughly constant below $T_c$, then dropping 
smoothly to zero above. The strong $T$-dependence of $\Delta$ makes
it difficult to unambiguously determine $\Gamma_0$ from the
fits at this {\bf k}-point. On theoretical grounds, we expect that, here too, 
there is a non-zero $\Gamma_0$, and the closing of the pseudogap
is again determined by $\Delta(T) \sim \Gamma_0(T)$, however this condition
is satified by the rapid drop in $\Delta(T)$, rather than the rise in 
$\Gamma_0(T)$.  For completeness, we also show $\Delta(T)$ for this sample
at the $(\pi,0)-(\pi,\pi)$ Fermi crossing, which has a similar behavior to
that of the 83K sample.

We see that these results give further evidence for the unusual
{\bf k}-dependences first noted in ref.~\onlinecite{NEW}.
Strong pairing correlations are seen over a very wide $T$-range
near $(\pi,0)$, but these effects are less pronounced
and persist over a smaller $T$-range as one moves closer to the 
diagonal. This is clearly tied to the strong
{\bf k}-dependence of the effective interaction and the unusual
(anomalously broad and non-dispersive) nature of electronic states 
near $(\pi,0)$. Some of these features are captured in recent
theoretical studies of the pseudogap \cite{GESH,WEN,JAN}.

To summarize, we have introduced a simple phenomenological
self-energy expression which helps us to analyze ARPES data
and gain insight into the $T$-dependences of important parameters
like the gap and the one-particle scattering rate. Perhaps the most
interesting new result concerns the modeling of the pseudogap
data in the underdoped cuprates, where we found a
new lifetime effect above $T_c$ proportional to $T-T_c$. We argue that
this is naturally explained as an effect of pairing correlations above
$T_c$ on the one-particle spectral function. 
In fact, this qualitative observation argues
against non-pairing theories of the pseudogap, in that a
term proportional to $T-T_c$ does not naturally appear in such theories.
A second important result concerns the differences between the
pseudogap behavior in different parts of {\bf k}-space.
Near $(\pi,0)$ we found a constant $\Delta(T)$ with ``filling in'' of 
spectral weight with increasing $T$.  Away from this region, 
the pseudogap closes rather than fills in. 
These observations impose very important constraints for the
microscopic theory of high $T_c$ cuprates.

{\bf Appendix:} 
The simplest way to motivate the second term of Eq.~2 
is as follows.  The BCS $\Sigma$ of Eq.~1 may be seen
as arising from a bare Green's function $G_0$ dressed
by a pair ``susceptibility'' of the form 
$\Delta_{\bf k}^2\delta({\bf q})\delta(\omega)$
representing static long range order.  If the
pairs have a finite inverse lifetime $\Gamma_0$, then $\delta(\omega)$
is broadened into a Lorentzian, leading to Eq.~2. 

We next sketch a more formal derivation, valid in the regime of
``small fluctuations''\cite{FLUCT}, which gives further insight 
into the form of of $\Gamma_0(T)$. 
The lowest order graph is $G_0$ dressed by the fluctuation pair 
propagator $L$:
$\Sigma({\bf k},\epsilon_n) = 
-T \sum_\nu \int (d{\bf q}) L({\bf q},\omega_\nu)
G_0({\bf q}-{\bf k},\omega_\nu -\epsilon_n)$.
Here $L^{-1}({\bf q},\omega_\nu) =  
N_0(\varepsilon + \alpha|\omega_\nu| + q^2\xi_0^2)$
with $\varepsilon \simeq (T- T_c)/T_c$, 
$\alpha = \pi/8T_c$, the coherence length $\xi_0 \sim v_F/T_c$, 
and $N_0$ is the density of states. 
$\epsilon_n$ and $\omega_\nu$ are Fermi and Bose frequencies
respectively and $\int (d{\bf q}) = \int d^D q/(2\pi)^D$.
Evaluating $\sum_\nu$ using standard contour integration, we obtain
$\Sigma({\bf k},\epsilon_n) = (T/N_0)\int (d{\bf q}) 
F_1({\bf q})F_2({\bf q})$.
Here $F_1({\bf q}) = 1/[\varepsilon + q^2\xi_0^2]$
is sharply peaked in $q$ with a scale $\xi_0^{-1} \sqrt{\varepsilon}$. 
Before analytic continuation $|i\epsilon_n|  \ge \pi T$, and thus
the $q$-variation of
$F_2({\bf q}) = 1/[i\epsilon_n + i8T_c(\varepsilon + q^2\xi_0^2)/\pi
+ \epsilon({\bf q} - {\bf k})]$ is on the much larger scale of $\xi_0^{-1}$.
Thus we make the approximation
$\Sigma \simeq F_2(0) (T / N_0) \int d({\bf q})F_1({\bf q})$. 
Finally, doing the Bose sum in $\langle |\Delta({\bf r},t)|^2 \rangle
= T \sum_\nu \int (d{\bf q}) L({\bf q},\omega_\nu)$, 
we get the ``fluctuating'' gap
$\langle |\Delta|^2 \rangle = (T/N_0) \int(d{\bf q})F_1({\bf q})$. 
Substituting this in $\Sigma$, and setting 
$i\epsilon_n \rightarrow \omega + i0^+$, we obtain the second term in
(\ref{SE.2}) with $\Gamma_0(T) = 8(T - T_c)/\pi$.

In $D=0$ dimensions, which may be relevant in the vicinity of the
dispersionless $(\pi,0)$-point, we do not need to
make the approximation of pulling $F_2(0)$ out of the $q$-integration,
which justifies the result even for arbitrarily small $\omega$.
In any case, we emphasize that above derivation is used here only to motivate
the form of Eq.~2, which is then
used for fits in a more general context.

{\bf Acknowledgements;}
We thank Yuri Vilk, Andrei Varlamov, Boldizar Janko, and Jim Allen for useful
discussions.
This work was supported by the U. S. Dept. of Energy,
Basic Energy Sciences, under contract W-31-109-ENG-38, the National 
Science Foundation DMR 9624048, and
DMR 91-20000 through the Science and Technology Center for
Superconductivity.

\begin{figure}
\epsfxsize=3.2in
\epsfbox{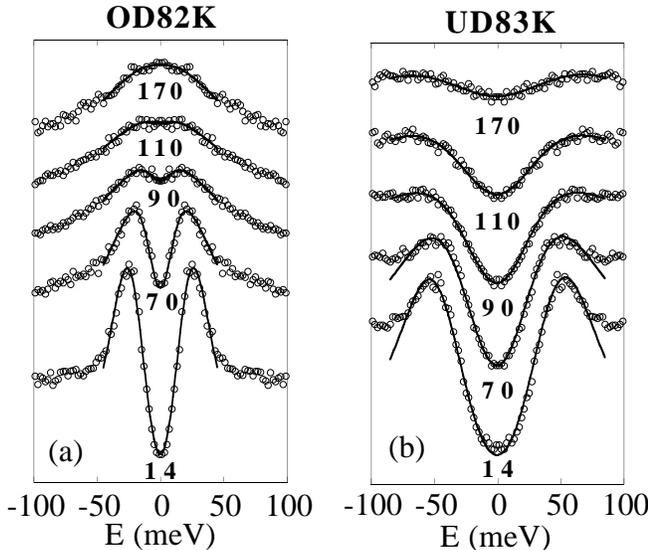}
\vspace{0.5cm}
\caption{
Symmetrized data for (a) a $T_c$=82K overdoped sample and (b) a $T_c$=83K 
underdoped sample at the $(\pi,0)-(\pi,\pi)$ Fermi
crossing at five temperatures, compared to
the model fits described in the text.}
\label{fig1}
\end{figure}

\begin{figure}
\epsfxsize=3.2in
\epsfbox{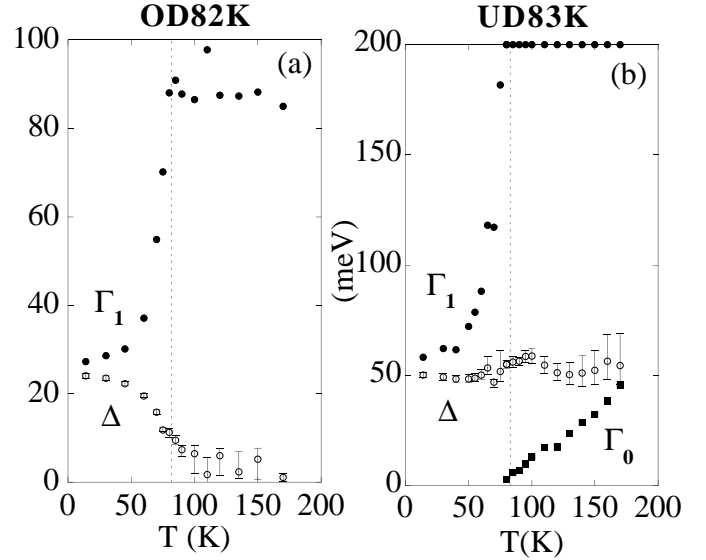}
\vspace{0.5cm}
\caption{
$\Delta$ (open circles), $\Gamma_1$ (solid circles), and $\Gamma_0$ (solid
squares) versus $T$ at the $(\pi,0)-(\pi,\pi)$ Fermi crossing for (a) a
$T_c$=82K overdoped sample and (b) a $T_c$=83K underdoped sample.
 The dashed line marks $T_c$.  The error bars for
$\Delta$ are based on a 10\% increase in the RMS error of the fits.}
\label{fig2}
\end{figure}

\begin{figure}
\epsfxsize=3.2in
\epsfbox{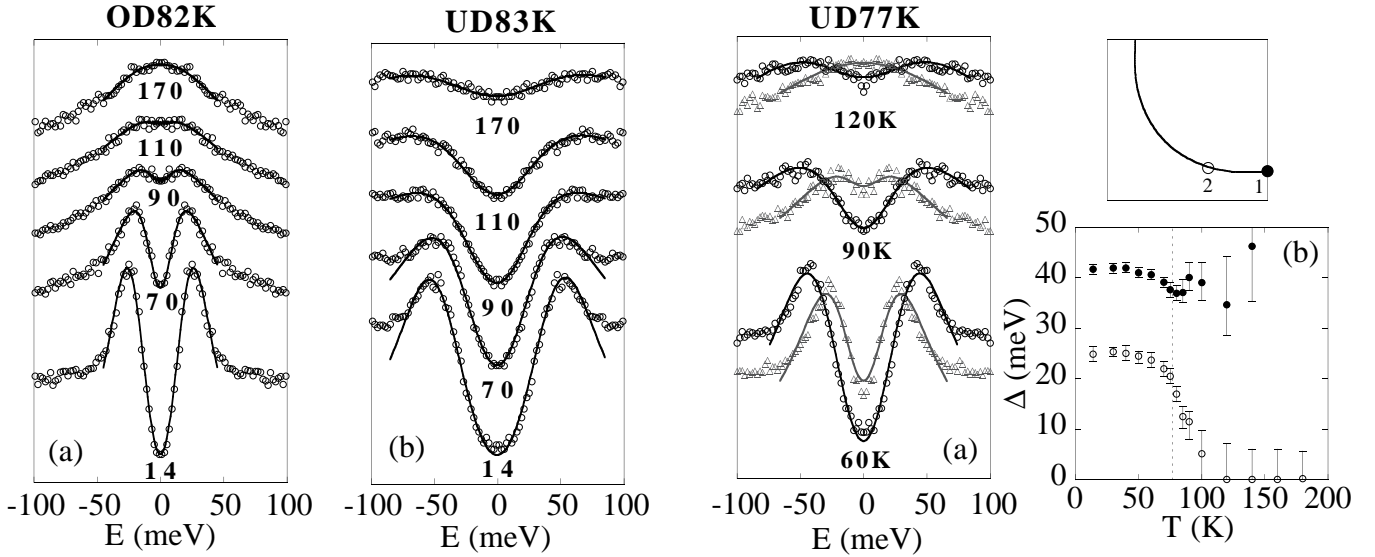}
\vspace{0.5cm}
\caption{
(a) Symmetrized data for a $T_c$=77K underdoped sample for three temperatures
at (open circles) ${\bf k}_F$ point 1 in the zone inset, and at (open
triangles) ${\bf k}_F$ point 2, compared to the model fits.
(b) $\Delta(T)$ for these two k points (filled
and open circles), with $T_c$ marked by the dashed line.}
\label{fig3}
\end{figure}


\begin{references}

\bibitem{NK}
M. Randeria {\it et al.}, Phys. Rev. Lett. {\bf 74}, 4951 (1995).

\bibitem{STANFORD}
D. S. Marshall {\it et al.}, Phys. Rev. Lett. {\bf 76}, 4841 (1996);
A. G. Loeser {\it et al.}, Science {\bf 273}, 325 (1996).

\bibitem{DING}
H. Ding {\it et al.}, Nature {\bf 382}, 51 (1996).

\bibitem{PRL97}
H. Ding {\it et al.}, Phys. Rev. Lett. {\bf 78}, 2628 (1997).

\bibitem{REVIEW}
For a detailed review and references, see: M. Randeria,
Varenna Lectures, 1997; cond-mat/9710223.

\bibitem{NEW}
M. R. Norman {\it et al.}, preprint (1997); cond-mat/9710163

\bibitem{FOOT1}
Spectra analyzed below were normalized to the constant 
signal above the Fermi energy, which was then subtracted off.
No other background subtractions were made since we are focussing
on low energy data.

\bibitem{HIGH}
The deviations beyond about 40 meV at low $T$
are due to the strong $\omega$-dependence of the SC state self-energy
for $\omega > 2\Delta$;
see M. R. Norman {\it et al.}, Phys. Rev. Lett. {\bf 79}, 3506 (1997).

\bibitem{PUCH}
A. V. Puchkov, D. N. Basov, and T. Timusk, J. Phys. Cond. Matter
{\bf 8}, 10049 (1996).

\bibitem{RESIDUAL}
The residual term $a$ is due to the finite $\omega$ range of the fit.
Reducing this range leads to a reduction in $a$.  

\bibitem{TRANSPORT}
D. A. Bonn, P. Dosanjh, R. Liang, and W. N. Hardy, Phys. Rev. Lett.
{\bf 68}, 2390 (1992);
K. Krishana, J. M. Harris, and N. P. Ong, Phys. Rev. Lett. {\bf 75},
3529 (1995).

\bibitem{LORAM}
J. W. Loram {\it et al.}, J. Supercond. {\bf 7}, 243 (1994).

\bibitem{RENNER}
Ch. Renner {\it et al.}, submitted, Phys. Rev. Lett.

\bibitem{HARRIS}
J. M. Harris {\it et al.}, Phys. Rev. B {\bf 54}, R15665 (1996).

\bibitem{SNS97}
H. Ding {\it et al.}, SNS97 proceedings, to be published,
J. Phys. Chem. Solids.

\bibitem{MIYA}
N. Miyakawa {\it et al.}, to be published, Phys. Rev. Lett.

\bibitem{GESH}
V. B. Geshkenbein, L. B. Ioffe, and A. I. Larkin, Phys. Rev. B {\bf 55},
3173 (1997).

\bibitem{WEN}
X. G. Wen and P. A. Lee, Phys. Rev. Lett. {\bf 76}, 503 (1996).

\bibitem{JAN}
J. R. Engelbrecht, A. Nazarenko, M. Randeria, E. Dagotto,
preprint (1997); cond-mat/9705166.

\bibitem{FLUCT}
E. Abrahams, M. Redi and J. Woo,  Phys. Rev. B {\bf 1}, 208 (1970);
A. Schmid, Z. Physik {\bf 231}, 324 (1970);
B. R. Patton, PhD thesis, Cornell Univ. (1971).

\end{references}
\end{document}